# Quantum Controlled Error-Correction Teleportation (QCECT) Scheme for Satellite Control and Communication

William Thacker, PhD | Aakash Gupta
Saint Louis University, MO

**Abstract.** A 5-qubit quantum controlled teleportation scheme incorporated with a 3-qubit quantum error-corrective circuit is proposed for satellite control and communication. This means that the coordinates of a satellite on the unit two-dimensional Bloch sphere representation can be sent to the ground based operator via quantum teleportation. Applying quantum rotation operators to the initial quantum state, $|\psi\rangle$, to obtain desired changes, the corrected/desired quantum state, $|\psi'\rangle$, can be sent back to the satellite to adjust its coordinates.

**Keywords.** Quantum Teleportation | Quantum Error-Correction | Bloch Sphere | Quantum Rotation Operators | Quantum Measurement

## 1 Introduction

Quantum Teleportation has always been an interesting outcome of Quantum Communication and Quantum Information theory that allows transportation of an unknown quantum state from a sender to a spatially distant receiver with the help of shared e-bits and classical communication. Several quantum teleportation protocols have been presented to achieve Alice-Bob's teleported communication. Several authors have contributed in enriching the database of quantum gates for teleporting quantum bits via special entangled states.[1-5] Duan, Zha, Sun, and Xia created a scheme for Bidirectional Quantum Controlled Teleportation (BQCT) via a maximally seven-qubit entangled state, where to-and-fro quantum communication can be made possible between Alice and Bob.[6]

In this paper, BQCT is developed upon 5-qubit quantum state, accomplished by an error-corrective scheme of 3-qubits for a noise tolerant communication between satellite and ground-based operator that is made possible for satellite control and communication. The algorithm of a quantum circuit developed encodes a 1-qubit quantum state with two ancillae along with a shared entangled EPR pair between the operator and the satellite. The quantum state $|\psi\rangle$ containing the instantaneous coordinates of the current orientation of the satellite in unit two-dimensional Bloch sphere passes through a quantum encryption circuit to process quantum information reliably in the presence of noise. The 3-qubit encoded quantum state along with the entangled EPR pair then goes through the quantum teleportation circuit, followed by syndrome diagnosis and recovery of the encoded qubit right before the measurements of the quantum teleportation protocol are taken. The outcomes of the measurement taken at the satellite are then sent to the ground-based operator via classical communication channel. Appropriate unitary transformation operators are made to act on the entangled qubit at the operator to decipher the quantum state $|\psi\rangle$ of the satellite containing its coordinates. The teleported state $|\psi\rangle$ then passes through Toffoli gate for central processing unit without affecting $|\psi\rangle$. Appropriate rotational and phase transformation operators are applied to the quantum state $|\psi\rangle$ to obtain the desired change in coordinates of the satellite, when needed, giving $|\psi'\rangle$. Similarly, the new quantum state $|\psi'\rangle$ is teleported to the satellite from the operator completing the quantum loop circuitry.

## 2 Quantum Controlled Error-Correction Teleportation (QCECT)

Quantum controlled error-correction teleportation scheme as proposed in this research can be achieved by the quantum circuit shown in Fig. 1 that incorporates a series of quantum gates achieving quantum encoding and quantum controlled teleportation of a quantum state $|\psi\rangle$ from the satellite to the ground-based operator, quantum rotation operators that act on $|\psi\rangle$ to give $|\psi'\rangle$, and teleportation of the desired quantum state $|\psi'\rangle$ back to the satellite as a control mechanism.

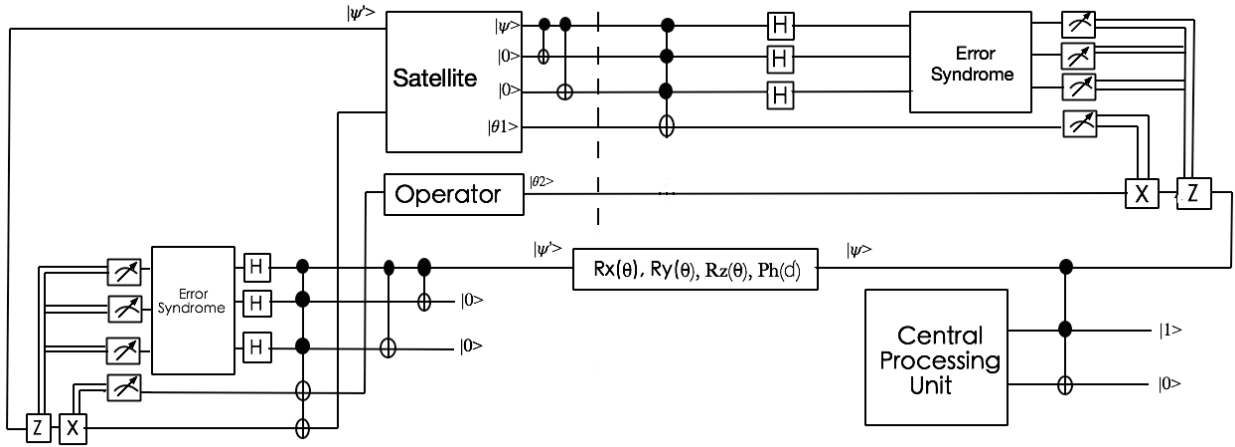

*Fig. 1: Quantum Controlled Error-Correction Teleportation (QCECT) Circuit*

The scheme can be described as follows. Suppose the coordinates of a satellite in a two-dimensional space can be represented through an arbitrary single qubit of a unit two-dimensional Bloch sphere in an unknown state, which is described by

$$|\psi_0\rangle = \cos\frac{\theta}{2}|0\rangle + e^{i\varphi}\sin\frac{\theta}{2}|1\rangle$$

(1)

The data to be encoded enters the encoding circuit for the three qubit bit flip code on the top channel with duo-ancillae on the bottom two channels, giving the encoded quantum state

$$|\psi_1\rangle = \cos\frac{\theta}{2}|000\rangle + e^{i\varphi}\sin\frac{\theta}{2}|111\rangle \tag{2}$$

At this stage, the encoded quantum state is ready to be teleported via quantum teleportation circuit. Equation 2 serves as the state input to the circuit. The EPR state that is assumed in this research is the Bell State $|\beta_{00}\rangle = \frac{1}{\sqrt{2}}(|00\rangle + |11\rangle)$. The first three channels comprise the encoded quantum data and the last two channels carry an EPR state shared between the satellite and the ground-based operator. Thus the full 5-qubit state at the beginning of the teleportation circuit is given by

$$|\psi_2\rangle = \frac{1}{\sqrt{2}}\left[\cos\frac{\theta}{2}|000\rangle(|00\rangle + |11\rangle) + e^{i\varphi}\sin\frac{\theta}{2}|111\rangle(|00\rangle + |11\rangle)\right] \tag{3}$$

The satellite sends it encoded qubits through a $C^3NOT$ gate, obtaining

$$|\psi_3\rangle = \frac{1}{\sqrt{2}}\left[\cos\frac{\theta}{2}|000\rangle(|00\rangle + |11\rangle) + e^{i\varphi}\sin\frac{\theta}{2}|111\rangle(|10\rangle + |01\rangle)\right] \tag{4}$$

The satellite then sends its encoded qubit through a Hadamard gate, obtaining

$$\begin{aligned}|\psi_4\rangle &= \frac{1}{4}\Big[\cos\frac{\theta}{2}(|0\rangle + |1\rangle)(|0\rangle + |1\rangle)(|0\rangle + |1\rangle)(|00\rangle + |11\rangle) \\
&\quad + e^{i\varphi}\sin\frac{\theta}{2}(|0\rangle - |1\rangle)(|0\rangle - |1\rangle)(|0\rangle - |1\rangle)(|10\rangle + |01\rangle)\Big] \\
&= \frac{1}{4}\Big[\cos\frac{\theta}{2}(|00\rangle + |01\rangle + |10\rangle + |11\rangle)(|0\rangle + |1\rangle)(|00\rangle + |11\rangle) \\
&\quad + e^{i\varphi}\sin\frac{\theta}{2}(|00\rangle - |01\rangle - |10\rangle + |11\rangle)(|0\rangle - |1\rangle)(|10\rangle + |01\rangle)\Big] \\
&= \frac{1}{4}\Big[\cos\frac{\theta}{2}(|000\rangle + |001\rangle + |010\rangle + |011\rangle + |100\rangle + |101\rangle + |110\rangle \\
&\quad + |111\rangle)(|00\rangle + |11\rangle) \\
&\quad + e^{i\varphi}\sin\frac{\theta}{2}(|000\rangle - |001\rangle - |010\rangle + |011\rangle - |100\rangle + |101\rangle + |110\rangle \\
&\quad - |111\rangle)(|10\rangle + |01\rangle)\Big]\end{aligned} \tag{5}$$

This is the stage where error detection takes place. A measurement is taken that identifies an error that might have occurred on the quantum state. According to the error syndrome registered, corresponding to the four possible projection operators as given below, the original encoded quantum state can be recovered accordingly.[1]

$P_0 \equiv |000\rangle\langle 000| + |111\rangle\langle 111|$ No Error
$P_1 \equiv |100\rangle\langle 100| + |011\rangle\langle 011|$ Bit flip on qubit one
$P_2 \equiv |010\rangle\langle 010| + |101\rangle\langle 101|$ Bit flip on qubit two
$P_3 \equiv |001\rangle\langle 001| + |110\rangle\langle 110|$ Bit flip on qubit three

According to the measurements result, the original quantum state can be recovered by applying a unitary bit flip operator to the respective affected qubit to give the original state.

Assuming that none of the qubits are corrupted, the quantum state $|\psi_4\rangle$ can be re-written in the following way, simply by regrouping terms

$$\begin{aligned}
|\psi_4\rangle = \frac{1}{4}\Big[ &|0000\rangle \left(\cos\frac{\theta}{2}|0\rangle + e^{i\varphi}\sin\frac{\theta}{2}|1\rangle\right) + |0001\rangle \left(\cos\frac{\theta}{2}|1\rangle + e^{i\varphi}\sin\frac{\theta}{2}|0\rangle\right) \\
+ &|0010\rangle \left(\cos\frac{\theta}{2}|0\rangle - e^{i\varphi}\sin\frac{\theta}{2}|1\rangle\right) + |0011\rangle \left(\cos\frac{\theta}{2}|1\rangle - e^{i\varphi}\sin\frac{\theta}{2}|0\rangle\right) \\
+ &|0100\rangle \left(\cos\frac{\theta}{2}|0\rangle - e^{i\varphi}\sin\frac{\theta}{2}|1\rangle\right) + |0101\rangle \left(\cos\frac{\theta}{2}|1\rangle - e^{i\varphi}\sin\frac{\theta}{2}|0\rangle\right) \\
+ &|0110\rangle \left(\cos\frac{\theta}{2}|0\rangle + e^{i\varphi}\sin\frac{\theta}{2}|1\rangle\right) + |0111\rangle \left(\cos\frac{\theta}{2}|1\rangle + e^{i\varphi}\sin\frac{\theta}{2}|0\rangle\right) \\
+ &|1000\rangle \left(\cos\frac{\theta}{2}|0\rangle - e^{i\varphi}\sin\frac{\theta}{2}|1\rangle\right) + |1001\rangle \left(\cos\frac{\theta}{2}|1\rangle - e^{i\varphi}\sin\frac{\theta}{2}|0\rangle\right) \\
+ &|1010\rangle \left(\cos\frac{\theta}{2}|0\rangle + e^{i\varphi}\sin\frac{\theta}{2}|1\rangle\right) + |1011\rangle \left(\cos\frac{\theta}{2}|1\rangle + e^{i\varphi}\sin\frac{\theta}{2}|0\rangle\right) \\
+ &|1100\rangle \left(\cos\frac{\theta}{2}|0\rangle + e^{i\varphi}\sin\frac{\theta}{2}|1\rangle\right) + |1101\rangle \left(\cos\frac{\theta}{2}|1\rangle + e^{i\varphi}\sin\frac{\theta}{2}|0\rangle\right) \\
+ &|1110\rangle \left(\cos\frac{\theta}{2}|0\rangle - e^{i\varphi}\sin\frac{\theta}{2}|1\rangle\right) + |1111\rangle \left(\cos\frac{\theta}{2}|1\rangle - e^{i\varphi}\sin\frac{\theta}{2}|0\rangle\right) \Big]
\end{aligned}$$
(6)

After the measurements are performed at the satellite on the first four qubit channels, the resulting classical information is needed to be sent to the ground-based operator via classical mode of communication.

For 0000, 0110, 1010, 1100 $\rightarrow |\psi_{5,1}\rangle = \left(\cos\frac{\theta}{2}|0\rangle + e^{i\varphi}\sin\frac{\theta}{2}|1\rangle\right)$

For 0001, 0111, 1011, 1101 $\rightarrow |\psi_{5,2}\rangle = \left(\cos\frac{\theta}{2}|1\rangle + e^{i\varphi}\sin\frac{\theta}{2}|0\rangle\right)$

For 0010, 0100, 1000, 1110 $\rightarrow |\psi_{5,3}\rangle = \left(\cos\frac{\theta}{2}|0\rangle - e^{i\varphi}\sin\frac{\theta}{2}|1\rangle\right)$

For 0011, 0101, 1001, 1111 $\rightarrow |\psi_{5,4}\rangle = \left(\cos\frac{\theta}{2}|1\rangle - e^{i\varphi}\sin\frac{\theta}{2}|0\rangle\right)$

(7)

The operator can then apply an appropriate unitary quantum gate to the qubit deciphered based on the classical information transmitted to obtain the original quantum state containing the coordinates of the satellite to be controlled.

For 0000, 0110, 1010, 1100 $\rightarrow Z^0X^0|\psi_{5,1}\rangle = I|\psi_{5,1}\rangle = |\psi_{5,1}\rangle$
For 0001, 0111, 1011, 1101 $\rightarrow Z^0X^1|\psi_{5,2}\rangle = aX|1\rangle + bX|0\rangle = |\psi_{5,1}\rangle$
For 0010, 0100, 1000, 1110 $\rightarrow Z^1X^0|\psi_{5,3}\rangle = aZ|0\rangle - bZ|1\rangle = a|0\rangle + b|1\rangle = |\psi_{5,1}\rangle$
For 0011, 0101, 1001, 1111 $\rightarrow Z^1X^1|\psi_{5,4}\rangle = Z(aX|1\rangle - bX|0\rangle) = aZ|0\rangle - bZ|1\rangle$
$$= a|0\rangle + b|1\rangle = |\psi_{5,1}\rangle$$
(8)

Thus, it can be noticed that the state $|\psi_{5,1}\rangle$ is the same as the original state $|\psi_0\rangle$ that the satellite was aiming to send to the operator.

Now, a Toffoli gate is applied to the deciphered original quantum state with $|1\rangle$ and $|0\rangle$ as ancillae on second and third quantum channel, respectively. This aids in processing the input data that is transmitted from the satellite at the central processing unit. Note, this does not affect the original quantum state $|\psi_5\rangle$ obtained.

Furthermore, the rotational operators can be made to act on the qubit $|\psi_5\rangle$ to induce a desired rotation of coordinates about x, y, and z axes giving $|\psi_6\rangle$ serving as $|\psi'\rangle$, as follows[1]

$$R_x(\theta) = e^{-i\theta\frac{X}{2}} = \cos\frac{\theta}{2}I - i\sin\frac{\theta}{2}X = \begin{bmatrix} \cos\frac{\theta}{2} & -i\sin\frac{\theta}{2} \\ -i\sin\frac{\theta}{2} & \cos\frac{\theta}{2} \end{bmatrix}$$

$$R_y(\theta) = e^{-i\theta\frac{Y}{2}} = \cos\frac{\theta}{2}I - i\sin\frac{\theta}{2}Y = \begin{bmatrix} \cos\frac{\theta}{2} & -\sin\frac{\theta}{2} \\ \sin\frac{\theta}{2} & \cos\frac{\theta}{2} \end{bmatrix}$$

$$R_z(\theta) = e^{-i\theta\frac{Z}{2}} = \cos\frac{\theta}{2}I - i\sin\frac{\theta}{2}Z = \begin{bmatrix} e^{-i\frac{\theta}{2}} & 0 \\ 0 & e^{i\frac{\theta}{2}} \end{bmatrix}$$

$$Ph(\delta) = e^{i\delta}\begin{bmatrix} 1 & 0 \\ 0 & 1 \end{bmatrix}$$

The modified quantum state $|\psi_6\rangle$ containing the desired satellite coordinates can then be teleported back to the satellite using the same scheme discussed in this paper.

## 3 Conclusion

Quantum controlled error-correction teleportation scheme serves as a reliable process for quantum information processing and quantum communication. The encoding circuit for the three qubit bit flip code makes the data transmission fault tolerant against any noise that might corrupt the initial state teleported to the ground-based operator.

A systematic scheme is given in this paper for error syndrome and the encoded quantum state recovery that will allow to successfully obtain a corrected 3-qubit encoded state. It turned out that the techniques of fault-tolerant quantum computation can be used to perform logical operations directly on encoded data, without the need to decode the data.[1] Hence, the corrected quantum state can be directly used for taking the measurements, skipping the decoding stage and thus reducing the processing time and possible errors.

The scheme utilizes simple unitary rotational operators that can be made to act on the original quantum state by the operator to achieve the desired coordinate of the satellite as an output.

The study is open to further implementations that can be made to enhance the quality and amount of information that can be teleported via QCECT by adding more quantum states to incorporate sophisticated data to be teleported, as needed.

**References**

1. Nielsen, M.A., Chuang, I.L. Quantum Computation and Quantum Information. Cambridge University Press, Cambridge (2002)

2. Gottesman, D., Chuang, I.L. (1999) Quantum Teleportation is a Universal Computational Primitive. Nature. DOI: 10.1038/46503

3. Shor, P. (1997) Fault-Tolerant Quantum Computation. AT&T Research, NJ. Retrived from: http://arxiv.org/pdf/quant-ph/9605011v2.pdf

4. Steane, A.M. (1998) Efficient Fault-Tolerant Quantum Computing. University of Oxford. Retrived from: http://arxiv.org/pdf/quant-ph/9809054v2.pdf

5. Brassard, G. (1996) Teleportation as a Quantum Computation. Physica D. DOI: 10.1016/S0167-2789(98)00043-8

6. Duan, Y., Zha, X., Sun, X., Xia, J. (2014) Bidirectional Quantum Controlled Teleportation via a Maximally Seven-qubit Entangled state. International Journal of Theoretical Physics. DOI: 10.1007/s10773-014-2065-1